\documentclass{article}

\usepackage{graphics}
\usepackage{amsmath,amssymb,mathabx}
\usepackage[utf8]{inputenc}
\usepackage[letterpaper,portrait,left=2.5cm,right=2.5cm]{geometry}

\begin{document}

\title{Dual characterization of critical fluctuations: Density functional theory \& nonlinear dynamics close to a tangent bifurcation}
\author{Mauricio Riquelme-Galván, Alberto Robledo \\ {\small Instituto de Física y Centro de Ciencias de la Complejidad, Universidad Nacional Autónoma de México} \\ {\small Apartado Postal 20-364, México 01000 DF, Mexico.}}

\date{}

\maketitle

\abstract{
We improve on the description of the relationship that exists between critical clusters in thermal systems and intermittency near the onset of chaos in low-dimensional systems. We make use of the statistical-mechanical language of inhomogeneous systems and of the renormalization group (RG) method in nonlinear dynamics to provide a more accurate, formal, approach to the subject. The description of this remarkable correspondence encompasses, on the one hand, the density functional formalism, where classical and quantum mechanical analogues match the procedure for one-dimensional clusters, and, on the other, the RG fixed-point map of functional compositions that captures the essential dynamical behavior. We provide details of how the above-referred theoretical approaches interrelate and discuss the implications of the correspondence between the high-dimensional (degrees of freedom) phenomenon and low-dimensional dynamics.
} 

\section{Introduction}
\label{intro}

Fluctuations are the dominant feature of criticality; their presence is spatially boundless while their duration is finite as they offer an ever-changing landscape. Their appearance is such that the critical state looks the same on any length scale. It is statistically scale invariant. Their ephemeral nature covers a range of life spans that manifest as a power spectrum of so-called $1/f$ noise \cite{Schuster}. We consider here a large critical fluctuation or cluster described by a shape of the order parameter $\phi$ in a region of size $R$.  Such critical clusters have been studied \cite{Antoniou1998,Antoniou2000,Contoyiannis2000,Contoyiannis2002,Robledo2005,Robledo2011} with the use of the statistical-mechanical method of a coarse-grained free energy, like the Landau-Ginzburg-Wilson (LGW) continuous spin model Hamiltonian, at the critical temperature and with zero external field. The statistically preponderant cluster of radius $R$ has been shown to have a fractal configuration \cite{Antoniou1998,Antoniou2000} whose amplitude $\phi$ grows in time and eventually collapses when an instability is reached \cite{Contoyiannis2000,Contoyiannis2002}. A nonlinear map describes this process with tangency and feedback features, such that the time evolution of the cluster is given in the nonlinear system as a laminar episode of intermittent dynamics \cite{Contoyiannis2000,Contoyiannis2002,Robledo2005,Robledo2011}.

Here we study such generic fluctuation or cluster of the order parameter $\phi$ for the particular case of a one-dimensional thermal system under a second order phase transition. The focus on one-dimensional systems provides access to a strait-forward spatial description of $\phi$ that captures some of the main properties of clusters in higher spatial dimensions, those that we are interested in highlighting here, while leaving others to be presented elsewhere. As known \cite{Fisher1972}, criticality in one-dimensional systems requires long range interactions. The dominant shape of the fluctuation or cluster in the large size $R$ limit is obtained via the saddle-point approximation on the partition function for the LGW Hamiltonian. This approach corresponds to the optimization of a free energy density functional for which we exhibit formal details. When calculating the first variation of the free energy functional the Euler-Lagrange (EL) equation obtained is analogous to that describing the motion of a particle in classical mechanics. From this analogy we determine the possible types of order parameter profiles for the critical cluster. We identify the force that the different profiles of the order parameter of the cluster exerts on the boundary and determines its growth or shrinkage. When calculating the second variation of the free energy functional an expression of the form of a Schrödinger equation for a quantum particle is obtained. The eigenfunctions of the Schrödinger equation correspond to perturbations of the order parameter profile while the eigenvalues indicate the stability of the profile to such perturbations.
 
After this, we focus on either the collapse or the intermittent behavior of the critical cluster. We transform the phase-portrait equation obtained (after one integration) from the EL equation into a nonlinear iteration map that is the starting point for the calculation of the renormalization group (RG) fixed-point map near tangency, as originally derived by Hu and Rudnick \cite{Rudnick1982}. Perturbation of this map \cite{Rudnick1982} takes it out of tangency and this provides, in one case, trajectories that describe the collapse of the cluster, and, in the other case, trajectories made of laminar episodes interrupted by rapid bursts, as is the intermittent dynamics associated to this type of nonlinear systems. The critical cluster develops during the laminar episode leading to sudden breakdown that is followed by the growth of a new cluster. Known \cite{Procaccia1983} power spectrum for this dynamical behavior delivers the $1/f$ noise characteristic of critical fluctuations. 

Finally, we comment on the link between the high-dimensional (many degrees of freedom) nature of systems at criticality and the simple low dimensional results for the dominant configurations that are represented by the saddle-point approximation on the coarse-grained free energy we used to describe the spatial and temporal features of a local fluctuation.

\section{Density functional theory for the shapes of critical clusters}
\label{sec:Densityfunctional}

We briefly recall the Landau two-stage approach to evaluate a partition function.  The partition function for the critical cluster is written as

\begin{equation}\label{eq:funcionparticion}
Z=\int_{\Omega}\mathcal{D}[\phi]Z_{\phi},
\end{equation}
where $Z_{\phi}=\exp(-\Gamma_{c}[\phi])$ is obtained by summing over the microscopic configurations that lead to a specific form for the order parameter $\phi(r)$ (where $r$ is the spatial position), $\Omega$ spans all $\phi$. The form of the exponent in Eq. \ref{eq:funcionparticion}, the Landau-Ginzburg-Wilson (LGW) effective Hamiltonian or free energy, is generally written as \cite{Antoniou1998,Antoniou2000}

\begin{equation} \label{eq:energiaLGW}
\Gamma_{c}[\phi]=a\int_{V} dx^{d}\left[ \frac{1}{2}(\nabla\phi)^{2}+b|\phi|^{\delta_{c}+1} \right],
\end{equation}

$V$ is the volume occupied by the cluster, $a$, $b$ are constants and $\delta_{c}$ is the isothermal critical exponent.

The sum over $Z_{\phi}$, actually a path integration, over all different $\phi(r)$ is replaced by a saddle-point approximation that delivers the particular form of $\phi(r)$ obtained from the dominant configurations. For simplicity we consider here a one-dimensional system, but the analysis can be extended to higher dimensions with similar results, albeit technically more complicated expressions \cite{Antoniou1998,Antoniou2000}. In dimension one Eq. \ref{eq:energiaLGW} reads

\begin{equation}\label{eq:Gammaunidimensional}
\Gamma_{c}[\phi]=a\int_{-R}^{R}dx\left[ \frac{1}{2}\left(\frac{d\phi}{dx}\right)^{2}+b|\phi|^{\delta_{c}+1} \right],
\end{equation}
where the size (length) of the cluster is $2R>0$. In one dimension the value of the exponent $\delta_{c}$ is determined by the decay rate of the interactions \cite{Fisher1972}.

\subsection{The first variation}
\label{subsec:firstvariation}
The saddle-point approximation for Eq. \ref{eq:funcionparticion} requires the vanishing of the first variation

\begin{equation}
\delta\Gamma_{c}[\phi]=a\left\{ \int_{-R}^{R}\left[ -\frac{\partial U}{\partial\phi}-\frac{d^2\phi}{dx^{2}} \right]\delta\phi dx+\left[ \frac{d\phi}{dx}\delta\phi \right]_{-R}^{R} \right\}=0,
\end{equation}
where $U(\phi)=-b|\phi|^{\delta_{c}+1}$. Since there are no boundary restrictions other than $\delta\phi(\pm R)=0$ the resulting Euler-Lagrange (EL) equation is simply

\begin{equation} \label{eq:EulerLagrange}
\frac{d^{2}\phi}{dx^{2}}=-\frac{dU}{d\phi}.
\end{equation}

\begin{figure}
\begin{center}
\resizebox{0.55\columnwidth}{!}{
  \includegraphics{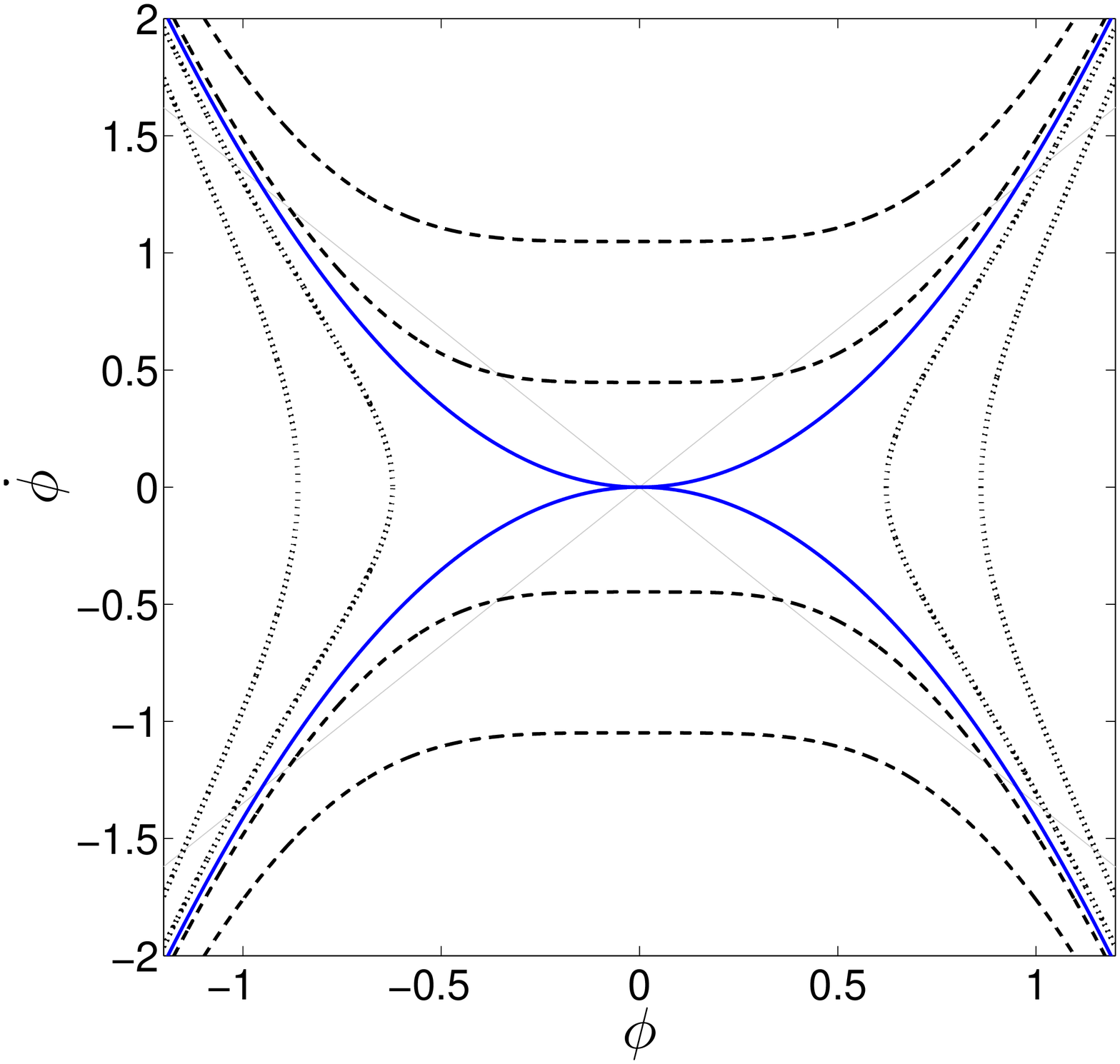}
}
\caption{Phase portrait $(\phi, \dot{\phi})$ for the classical value of $\delta_{c}=3$ and $b=1$. The solid lines show the separatrix $C _{\infty}=0$, the dotted lines at the left or at the right of the separatrix correspond to different values of $C_{R}<0$, the dashed lines above or below of the separatrix correspond to different values of $C_{R}>0$. The two straight lines crossing at $\phi, \dot{\phi}=0$ are used to introduce boundary conditions that stabilize the dominant clusters $\phi(x)$, $-R<x<R$.  See text for explanation.}
\label{fig:fase}
\end{center}
\end{figure}

Interestingly, Eq. \ref{eq:EulerLagrange} can be read as that for a classical-mechanical particle under potential $U(\phi)$.  Integration of the EL equation leads to

\begin{equation} \label{eq:CR}
C_{R}=\frac{1}{2}\left( \frac{d\phi}{dx} \right)^{2}-b|\phi|^{\delta_{c}+1},
\end{equation}

where the integration constant $C_{R}$ can be identified as the total energy of the particle. Cluster configurations with $C_{R}\neq 0$ contribute to the partition function Eq. \ref{eq:funcionparticion} proportionally to $\exp(-bRC_{R})$, so that the dominant saddle points are those for which $C_{R}\approx 0$. We can draw phase portraits $(\phi,\dot{\phi})$, see Fig. \ref{fig:fase}, by rewriting Eq. \ref{eq:CR} as

\begin{equation} \label{eq:phi'}
\dot{\phi}\equiv\frac{d\phi}{dx}=\pm\sqrt{2(C_{R}+b|\phi|^{\delta_{c}+1})},
\end{equation}

and then extract the dominant order-parameter profiles $\phi(x)$ via integration

\begin{equation} \label{eq:integral}
x=\text{sign}(\dot{\phi}) \int^{\phi(x)}_{\phi_{0}} \frac{d\phi}{\sqrt{2\left( C_{R}+b|\phi|^{\delta_{c}+1} \right)}}.
\end{equation}

For the inhomogeneous system of size $2R$ the constant of integration $C_{R}$ is an unbalanced force exerted by/on the cluster. The LGW free energy evaluated at the solution of the EL equation is written as

\begin{equation}\label{eq:Gammastationary}
\Gamma^{(s)}_{c}[\phi]=a\int^{\phi(R)}_{\phi(-R)} \dot{\phi}d\phi - 2aC_{R}R,
\end{equation}

so that

\begin{equation} \label{eq:CRg}
C_{R}=-\frac{1}{2a}\frac{\partial \Gamma^{(s)}_{c}[\phi]}{\partial R}.
\end{equation}

Therefore, $C_{R}$ can be associated with the forces exerted at the cluster boundaries.
To balance the forces (and convert the fluctuation into a stationary system) boundary conditions need to be added to $\Gamma_{c}[\phi]$ in Eq. \ref{eq:Gammaunidimensional}. These are, for example, $\Phi(\pm R)=1/2g\phi^{2}(\pm R)$, where the quantity $g$ is the so-called surface enhancement parameter \cite{Binder1983}. These boundary conditions appear as the straight lines with slopes $\pm g$ in Fig. \ref{fig:fase}. 

\begin{figure}
\begin{center}
\resizebox{0.55\columnwidth}{!}{
  \includegraphics{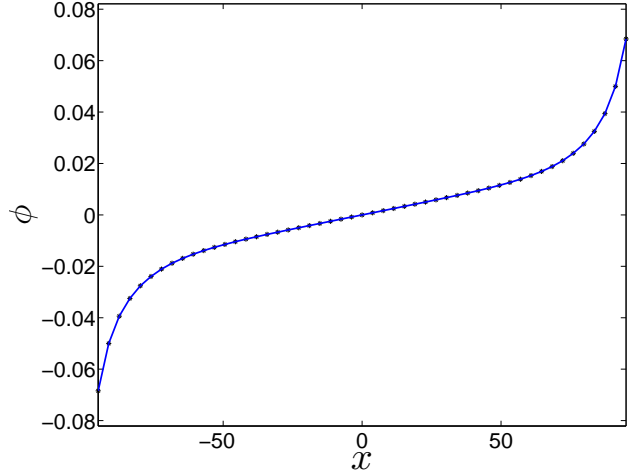}
}
\caption{Antisymmetric cluster order parameter profile $\phi(x)$ as obtained from the appropriate branch of the phase portrait in Fig. \ref{fig:fase}. The cluster is under an unbalanced force, $C_{R}=2.4\times 10^{-9}$, that indicates it should expand until the encounter of an instability. See text for description.}
\label{fig:perCpos}
\end{center}
\end{figure}

\begin{figure}
\begin{center}
\resizebox{0.55\columnwidth}{!}{
  \includegraphics{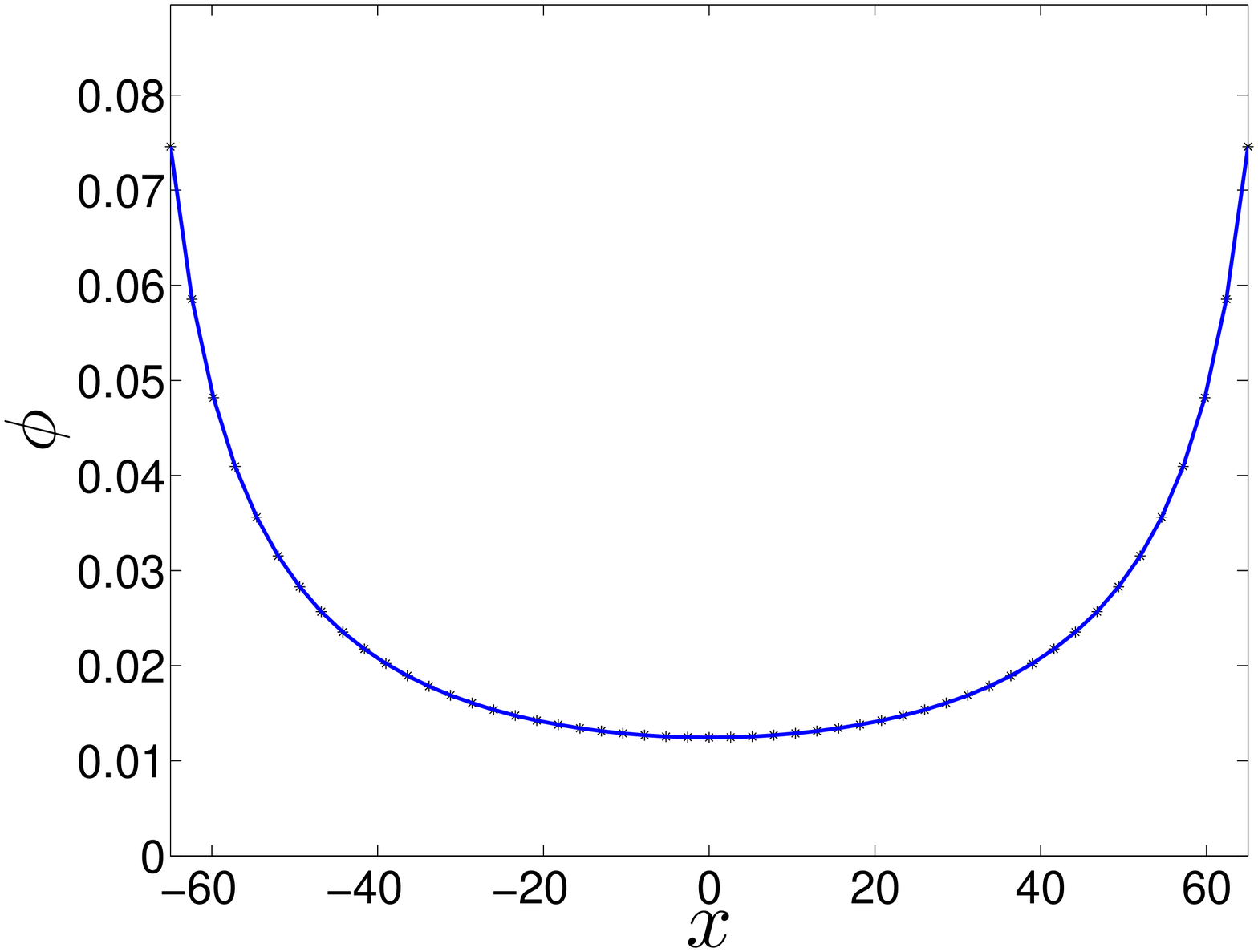}
}
\caption{Symmetric cluster order parameter profile $\phi(x)$ as obtained from the appropriate branch of the phase portrait in Fig. \ref{fig:fase}. The cluster is under an unbalance force, $C_{R}=-2.4\times 10^{-9}$, that indicates it should shrink until disappearance. See text for description.}
\label{fig:perCneg}
\end{center}
\end{figure}

We have the following two cases. First, the profile $\phi(x)$ is symmetric with respect to $x=0$, it is extracted from a branch of the phase portrait at the left or at the right of the separatrix. See dotted curves in Fig. \ref{fig:fase}. The profile is obtained from integration of the phase portrait branch traversing from its intersection with the line of slope $-g$ to the line of slope $g$. The force acting on this profile is $C_{R}<0$ and therefore the cluster shrinks in size until it disappears. Second, the profile $\phi(x)$ is antisymmetric with respect to $x=0$, it is obtained from a branch of the phase portrait above or below the separatrix . See dashed curves in Fig. \ref{fig:fase}. Again, the profile is determined from integration of the phase portrait branch crossing from its intersection with the line of slope $-g$ to the line of slope $g$. The force acting on this profile is $C_{R}>0$ and therefore the cluster expands in size until it encounters an instability (described in Section \ref{sec:RenormalizationGroup}). In Figs. \ref{fig:perCpos} and \ref{fig:perCneg} we show the antisymmetric and symmetric cluster profiles, respectively.

In terms of the classical-mechanical analog the profiles are particle trajectories with a time of flight equal to $2R$. The boundary conditions act as take-in or take-out walls that restore the energy balance and keep the particle repeating its movement.

When the constant of integration vanishes the order-parameter profile $\phi(x)$ can be determined analytically in closed form

\begin{equation} \label{eq:phiCcerodeformado}
\phi(x)=\phi_{0}\exp_{q}\left[|\phi_{0}|^{q-1}ux \right],
\end{equation}

where $\exp_{q}$ denotes the $q$-exponential function, $\exp_{q}(x)\equiv [1-(q-1)x]^{-1/(q-1)}$, and where $u=\sqrt{2b}$, $q=(\delta_{c}+1)/2$, and $\phi_{0}=\phi(0)$ \cite{Robledo2005,Robledo2011}. This case corresponds to a cluster of infinite size $R\rightarrow\infty$, $C_{\infty}=0$.

\subsection{The second variation}
\label{subsec:SecondVariation}

The stability of the EL `stationary' profile $\phi_{s}(x)$ (for clarity we denote in this subsection the solutions $\phi(x)$ of Eq. \ref{eq:EulerLagrange} with the subindex $s$) of the critical cluster can also be determined from the sign of the second variation

\begin{equation}\label{eq:secondvariation}
\delta^{2}\Gamma_{c}=a\int_{-R}^{R}\left[ -\left.\frac{\partial^{2}U}{\partial\phi^{2}}\right|_{\phi_{s}(x)}\delta\phi(x)-\frac{d^2\delta\phi(x)}{dx^2} \right]\delta\phi(x)dx,
\end{equation}
where $\delta\phi(x)\equiv\phi(x)-\phi_{s}(x).$

This time we have a description equivalent to that of a quantum-mechanical particle. Any perturbation $\delta\phi(x)$ of the order parameter can be decomposed as

\begin{equation}
\delta\phi(x)=\sum_{n}C_{n}\psi_{n}(x),
\end{equation}

where the $\psi_{n}(x)$ is a complete set of functions. In terms of these functions Eq. \ref{eq:secondvariation} can be written as \cite{Robledo1998}

\begin{equation}
\delta^{2}\Gamma_{c}[\phi]=\sum_{n}E_{n}(C_{n}\psi_{n}(x))^{2},
\end{equation}

where

\begin{equation} \label{eq:Schrodinger}
-\frac{d^{2}\psi_{n}(x)}{dx^{2}}+V(x)\psi_{n}(x)=E_{n}\psi_{n}(x),
\end{equation} 
is a single-particle Schrodinger equation under a potential

\begin{equation} \label{eq:potV}
V(x)=-\left.\frac{\partial^{2}U}{\partial\phi^{2}}\right|_{\phi_{s}(x)}.
\end{equation}
Therefore the sign of the second variation is given by the signs of the eigenvalues $E_{n}$ associated with the eigenfunctions $\psi_{n}(x)$, each of which represents a particular kind of variation $\delta\phi(x)$.

\begin{figure}
\begin{center}
\resizebox{0.55\columnwidth}{!}{
  \includegraphics{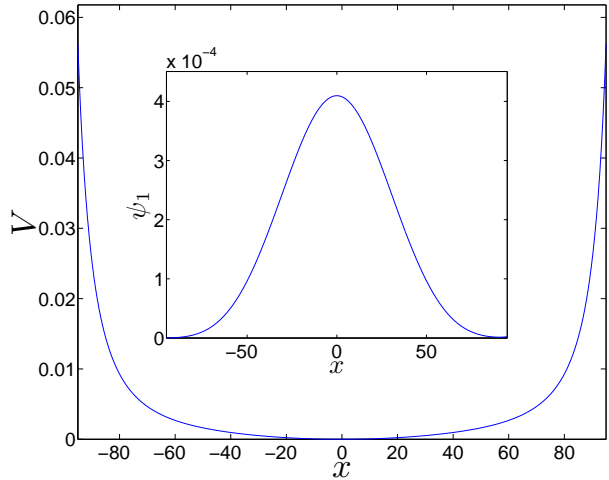}
}
\caption{The potential $V(x)$ in Eq. \ref{eq:potV} for the antisymmetric profile in Fig. \ref{fig:perCpos}. The inset shows the first eigenfunction. The first eigenvalue is $E_{1}=0.55\times 10^{-3}$ for $\delta_{c}=3$, $b=1$ and $C_{R}=2.4\times 10^{-9}$.}
\label{fig:potpert} 
\end{center}
\end{figure}

\begin{figure}
\begin{center}
\resizebox{0.55\columnwidth}{!}{
  \includegraphics{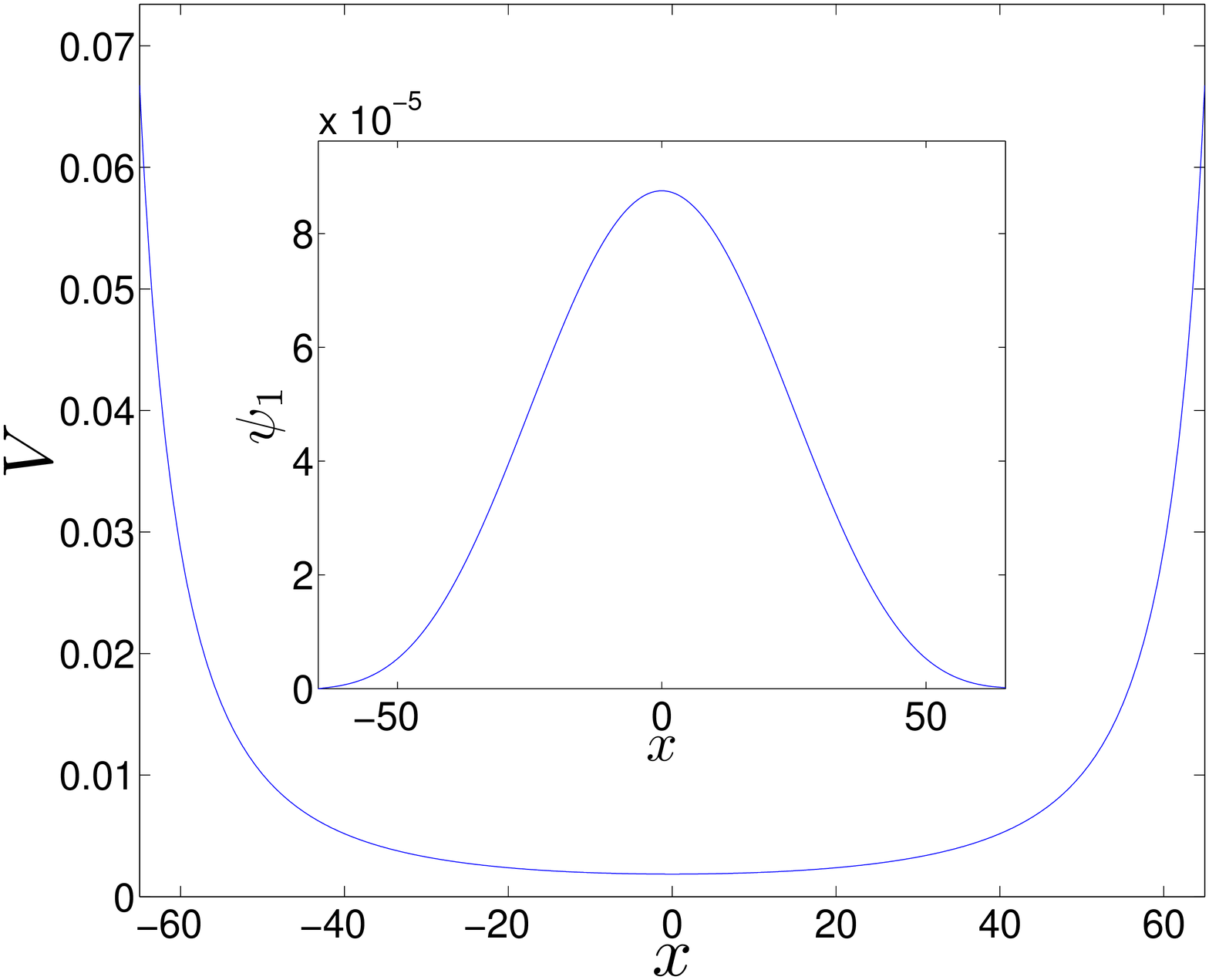}
}
\caption{The potential $V(x)$ in Eq. \ref{eq:potV} for the symmetric profile in Fig. \ref{fig:perCneg}. The inset shows the first eigenfunction. The first eigenvalue is $E_{1}=2.7\times 10^{-3}$ for $\delta_{c}=3$, $b=1$ and $C_{R}=-2.4\times 10^{-9}$.}
\label{fig:perpertneg}
\end{center}
\end{figure}

As before, the stability of the cluster profiles $\phi_s(x)$ is ensured by boundary conditions such as $\Phi(-R)$ and $\Phi(R)$ incorporated above into $\Gamma_{c}[\phi]$. These translate as boundary conditions when solving Eq. \ref{eq:Schrodinger}. In Figs. \ref{fig:potpert} and \ref{fig:perpertneg} we show, respectively, the potential $V$ and the first eigenfunctions calculated for the order parameter profiles $\phi_s(x)$ in Figs. \ref{fig:perCpos} and \ref{fig:perCneg} that are representative of the expanding antisymmetric and contracting symmetric cases described previously. We note that the expanding antisymmetric profile can be perturbed repeatedly by the first eigenfunction deformation with vanishing free energy cost as the vanishing of $V(0)$ at $x=0$ implies vanishing first eigenvalue for increasingly large $R$ (see \cite{Robledo1999}). Contrary to this, the positive value of $V(0)$ for the contracting symmetric profile implies a finite cost to all deformations.

\section{Renormalization Group fixed-point description of intermittency of critical clusters}
\label{sec:RenormalizationGroup}

We can transform the spatial description of the order parameter $\phi(x)$ into (a type of) time evolution of the cluster via a simple modification of the phase portrait Eq. \ref{eq:phi'}.  This is done replacing the derivative $d\phi/dx$ by the finite difference $(\phi_{t+1}-\phi_t)/((t+1)-t)$, so that we have

\begin{equation} \label{eq:mapaphi}
\phi_{t+1}=\phi_{t} \pm\sqrt{2(C_{R}+b|\phi_{t}|^{\delta_{c}+1})}.
\end{equation}

\begin{figure}
\begin{center}
\resizebox{0.55\columnwidth}{!}{
  \includegraphics{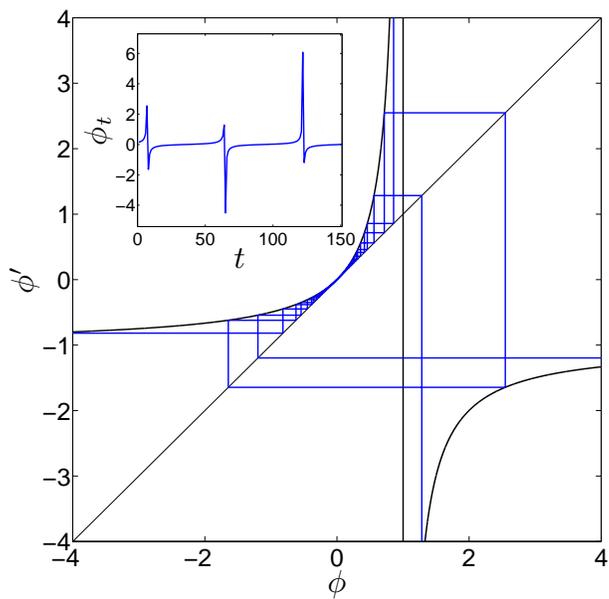}
}
\caption{Perturbed RG fixed-point map with $\epsilon<0$. A chaotic trajectory is shown performing passages through the map narrow channel close to the identity line mediated by reinjections that make use of the bottom branch of the map that occurs after the map singularity at $\phi_0$.}
\label{fig:mapadinamica}
\end{center}
\end{figure}

\begin{figure}
\begin{center}
\resizebox{0.55\columnwidth}{!}{
  \includegraphics{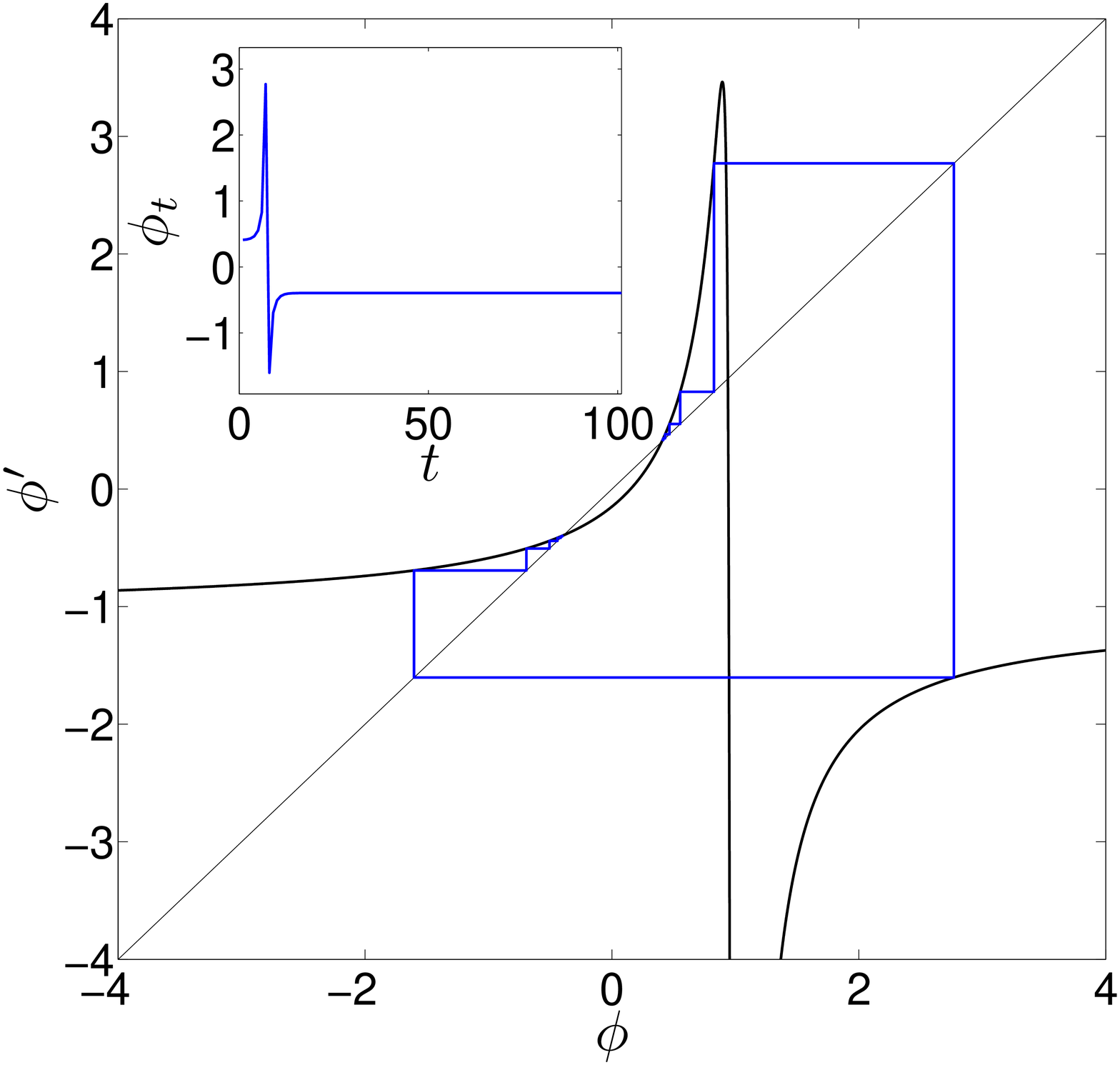}
}
\caption{Perturbed RG fixed-point map with $\epsilon>0$. The intersections of the map with the identity line, attractor and repellor positions, generate trajectories that are guided towards the attractor or driven away from the repellor.}
\label{fig:mapadinamica2}       
\end{center}
\end{figure}

\begin{figure}
\begin{center}
\resizebox{0.55\columnwidth}{!}{
  \includegraphics{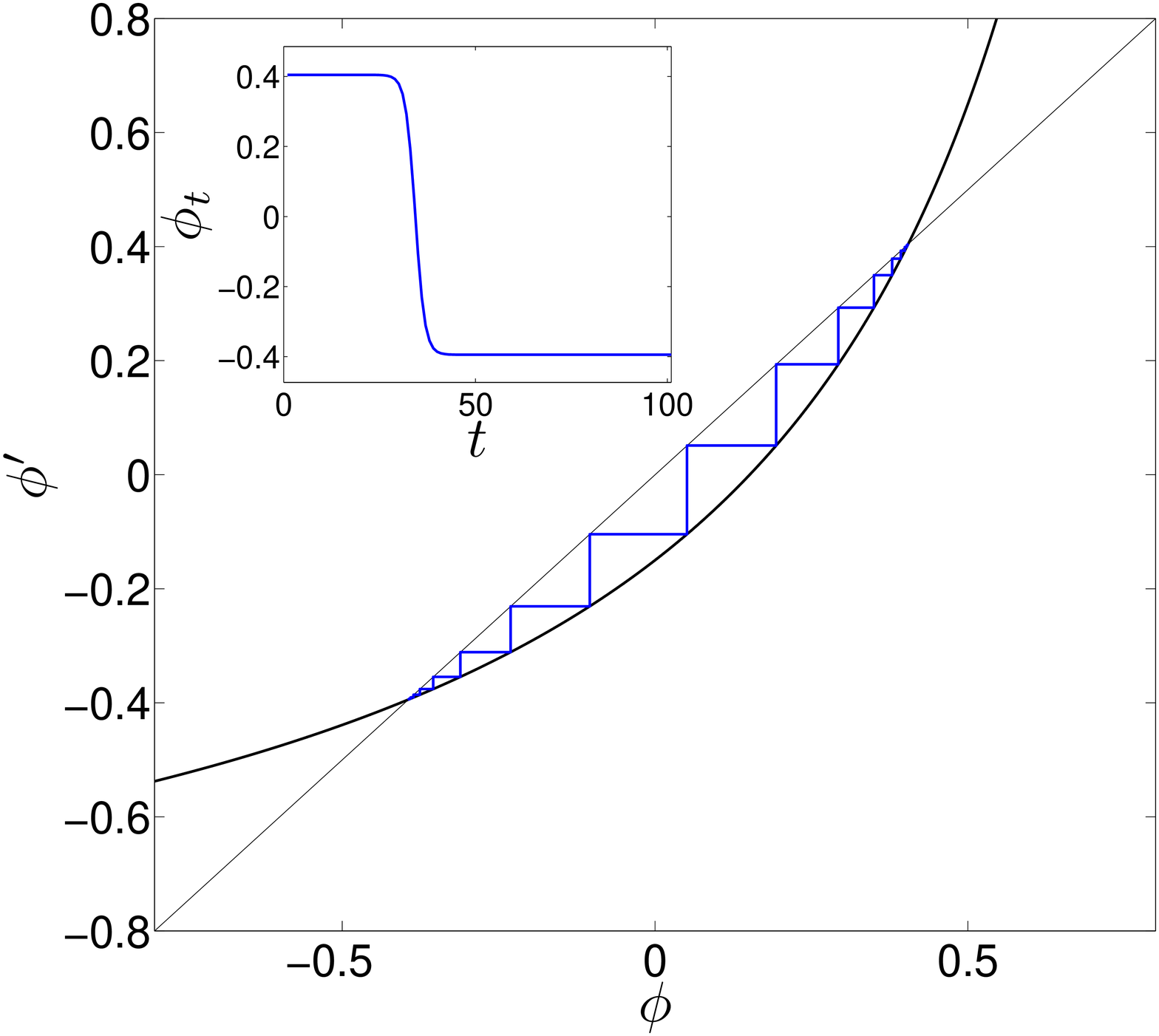}
}
\caption{Perturbed RG fixed-point map with $\epsilon>0$.Another trajectory that moves away from the repellor and converges to the attractor position.}
\label{fig:contraction}     
\end{center}
\end{figure}

Or, when $C_{R\rightarrow\infty}=0$

\begin{equation} \label{eq:mapaphiCcero}
\phi_{t+1}=\phi_{t} + \sqrt{2b}|\phi_{t}|^{\frac{\delta_{c}+1}{2}},
\end{equation}

a map equivalent to the local expression that is the starting point of the classic RG procedure to obtain the fixed-point map associated with the tangent bifurcation \cite{Schuster,Rudnick1982}. This is

\begin{equation} \label{eq:bifurcaciontangente}
\phi'=\phi+u\phi^{z}+O(|\phi|^{z}), \quad z>1, \quad u>0,
\end{equation}

where $u=\sqrt{2b}$ and $z=(\delta_{c} +1)/2$. An important property of the fixed-point map is that all the trajectories follow the closed form \cite{Robledo2002}

\begin{equation}
\phi_{t}=\phi_{0}\exp_{z}\left[ \phi_{0}^{z-1}ut \right],
\end{equation}

that, interestingly, can be compared with Eq. \ref{eq:phiCcerodeformado}. A perturbation takes the RG fixed-point map \cite{Rudnick1982} out of tangency

\begin{align} \label{eq:mapadinamica}
\phi_{t+1}=&\phi_{t}\exp_{z}\left[ u\phi_{t}^{z-1} \right] \notag \\
& +\frac{\epsilon}{z-1}\phi_{t}^{-(z-1)}\left( 1-(z-1)u\phi_{t}^{z-1}-\exp_{z}\left[ u\phi_{t}^{z-1} \right]^{z} \right) \\
& +O(\epsilon^{2}), \notag
\end{align}

where $z=(\delta_{c}+1)/2$ and $\epsilon$ is the perturbation parameter. Fig. \ref{fig:mapadinamica} shows the perturbed RG map when $\epsilon<0$ and corresponding trajectories with reinjection that produce successive laminar episodes separated by short chaotic bursts. Figs. \ref{fig:mapadinamica2} and \ref{fig:contraction} show the perturbed RG map when $\epsilon>0$ and corresponding trajectories that converge to a point.

The dynamical behavior in Figs. \ref{fig:mapadinamica}, \ref{fig:mapadinamica2} and \ref{fig:contraction} is consistent with the stability found in the previous section for the profiles in Figs. \ref{fig:perCpos} and \ref{fig:perCneg}, respectively. Intermittency for $\epsilon<0$ is well-matched with the growth of the antisymmetric profile and the instability of the cluster is portrayed by the outflow of the trajectory from the map bottleneck and its reinjection. Whereas the capture of the trajectory by a single attractor position for $\epsilon>0$ is compatible with the cluster contraction and collapse of the symmetric profile.

\section{Summary and discussion}
\label{sec:Summaryanddiscussion}

We have evaluated existing studies \cite{Contoyiannis2000,Contoyiannis2002,Robledo2005,Robledo2011} of large one-dimensional critical clusters that determine the spatial and temporal properties of the order parameter that arise by considering dominant configurations as obtained from the saddle-point approximation of a partition function defined via a Landau-Ginsburg-Wilson effective Hamiltonian or free energy.  We have done so by revisiting the concepts and analysis in above-referred works, and by making use of the formal procedures expressed, first, by the density functional theory of inhomogeneous systems, to obtain the spatial description of the critical cluster; and then, by the RG functional composition treatment of the dynamics near a tangent bifurcation, to obtain a picture of the time evolution of the critical cluster.

With regards to the spatial structure of the order parameter we found that the profile $\phi(x)$ for a large cluster of size $2R$ can only be extracted from a phase portrait branch away from the separatrix (see Fig. \ref{fig:fase}). There are two types of clusters, the first is a symmetric profile shown in Fig. \ref{fig:perCneg} that experiences a negative force $C_{R}<0$ on its boundaries so that it should contract until it disappears. The second profile has a shape with order-parameter reflection symmetry (See Fig. \ref{fig:perCpos}) and it exerts a positive force $C_{R}>0$ to its boundaries, and for this reason grows (until, as we have seen when describing its time evolution via an iterated nonlinear map, collapses and subsequently leads to the growth of another cluster). The phase portrait branches used to extract the profiles $\phi(x)$ move closer to the separatrix as the cluster size $R$ grows. When $R\rightarrow\infty$ the separatrix  leads to a cluster profile with vanishing force $C_{\infty}=0$ and therefore a stable stationary infinite cluster with $\phi(x)=0$ for (almost all) $x$. With the enhancement boundary terms the cluster becomes an inhomogeneous equilibrium confined system. Without them we have information about its fate, growth or shrinkage. The stability of the critical cluster was also considered through the second variation of the density functional. 

Concerning the temporal evolution of the critical fluctuations we have elaborated on the nonlinear one-dimensional map close to a tangent bifurcation that has been developed to describe the growth and collapse of clusters in terms of intermittency \cite{Contoyiannis2000,Contoyiannis2002,Robledo2005,Robledo2011}. A simple rewriting of an expression for the spatial structure of the critical cluster into a recurrence expression for consecutive `snapshots' leads to a nonlinear local map that is the starting point of the known functional composition RG approach to the tangent bifurcation. Next, we have made use of the known \cite{Rudnick1982} perturbation of the RG fixed-point map to produce appropriate global map versions capable of describing different possible types of temporal evolution. In this case too, depending on the sign of the perturbation parameter $\epsilon$, we obtain either contraction and disappearance of the cluster or steady growth until the encounter of an instability produces collapse and renewal. It is fitting to recall here that the intermittency generated by the deterministic RG fixed-point map \cite{Procaccia1983} displays the $1/f$ noise characteristic of critical fluctuations.

We have advanced a modest, one dimensional, but comprehensive picture of the spatial and temporal properties of the statistically dominant critical clusters based on precise employment of the density functional theory for inhomogeneous systems combined with the nonlinear dynamical evolution of the RG fixed-point map and related small-perturbation maps. The combined approach outlines how a scale invariant system with high-dimensional configuration space can be described in coarse-grained scales by low-dimensional equivalent model systems.

\section*{Acknowledgments}
M. R-G. acknowledges support from Sistema Nacional de Investigadores (CONACyT, SEP). AR acknowledges support from DGAPA-UNAM-IN103814 and CONACyT-CB-2011-167978 (Mexican Agencies).

\end{document}